\documentclass{iau-JDSS}
\usepackage{graphicx}
\usepackage{natbib}

\newcommand{\hii}{{\sc Hii}}
\newcommand{\ha}{\ensuremath{\textrm{H}\alpha}}

\title[Warm ionized medium] 
{Modern view of the warm ionized medium}

\author[Hill et al]   
{A. Hill$^1$, R. Reynolds$^2$, L. Haffner$^2$, \& K. Wood$^3$, \& G. Madsen$^4$}

\affiliation{$^1$CSIRO Astronomy \& Space Science, Epping, NSW, Australia; {\tt alex.hill@csiro.au}\\[\affilskip]
$^2$Department of Astronomy, University of
Wisconsin-Madison, USA\\[\affilskip]
$^3$School of Physics and Astronomy, University of St Andrews, Scotland\\[\affilskip]
$^4$Sydney Institute for Astronomy, School of Physics, The University of Sydney, NSW, Australia}

\pubyear{2012}
\volume{Volume 15}  
\pagerange{??}
\date{?? and in revised form ??}
\jname{Highlights of Astronomy, Volume 14}
\editors{Ian F Corbett, ed.}
\begin{document}

\maketitle

\begin{abstract}
We review the observational evidence that the warm ionized medium (WIM) is a major and physically distinct component of the Galactic interstellar medium. Although up to $\sim 20 \%$ of the faint, high-latitude \ha\ emission in the Milky Way may be scattered light emitted in midplane \hii\ regions, recent scattered light models do not effectively challenge the well-established properties of the WIM.
\end{abstract}

The discovery of free-free absorption of the Galactic synchrotron background led \citet{Hoyle:1963ve} to propose the existence of an extended layer of diffuse ionized gas, now known in the Milky Way as the warm ionized medium (WIM). Further evidence for the WIM came from the dispersion of pulsar signals \citep{Guelin:1974vr}. These observations predicted diffuse optical H recombination line emission. A subsequent search using sensitive, high spectral resolution Fabry-Perot spectrometers found H$\alpha$ emission in agreement with these predictions and a filling factor of $\sim 30 \%$ \citep{Reynolds:1973jo}.

The WIM is now well-established as a major component of the interstellar medium of the Milky Way and other disk galaxies \citep[see review by][]{Haffner:2009ev}. It consists of a plasma with nearly fully ionized hydrogen which extends more than $1 \textrm{ kpc}$ from the midplane. The warm and cold neutral media are optically thick to ionizing photons, but only Lyman continuum radiation from hot stars has the energy required to maintain the observed surface recombination rate of the WIM. This challenge has been largely resolved by models in which supernova-driven turbulence and superbubble structures allow Lyman continuum photons to propagate from the midplane to heights $|z| \sim 1 \textrm{ kpc}$ and ionize the WIM \citep[and references therein]{Wood:2010dg}.

The discovery of \ha\ emission from the WIM led to the discovery of other faint nebular emission lines from this warm, widespread plasma. These lines distinguish the WIM from the \hii\ regions surrounding hot stars, as demonstrated in Figure~2 of \citet{Madsen:2006fw}: along a sightline which contains a classical (locally-ionized) \hii\ region and emission from the WIM at different velocities, the \ha\ emission is much brighter in the \hii\ region while the [{\sc Nii}]$\lambda 6584$ and [{\sc Sii}]$\lambda 6716$ (hereafter [{\sc Sii}]) emission are each brighter in the WIM. This trend is observed in a variety of environments, including at $v_{\mathrm{LSR}} = 0$ at high Galactic latitudes, in the Perseus Arm, in M33, and in the edge-on galaxies NGC~891 and NGC~55 at heights $|z| \gtrsim 0.5 \textrm{ kpc}$: the [{\sc Nii}]$\lambda 6584 / \ha$ and [{\sc Sii}]$/ \ha$ line ratios are enhanced in the WIM compared to \hii\ regions \citep{Rand:1990gn,Ferguson:1996eq,Haffner:1999hi,Hoopes:2003ec,Madsen:2006fw}.

In combination with observations of the [{\sc Nii}]$\lambda 5755 / $[{\sc Nii}]$\lambda 6584$ line ratio, a direct temperature diagnostic, these observations indicate that the temperature in the WIM is $\approx 9000 \textrm{ K}$ while the temperature in classical \hii\ regions is $\approx 6000 \textrm{ K}$ \citep{Reynolds:2001ds}. Simultaneously, weak [{\sc Oiii}]$ \lambda 5007$, strong  [{\sc Oii}]$\lambda 3726$, and weak He{\sc\ i} intensities relative to \ha\ indicate that the ionization state of both He and O are lower in the WIM than in classical \hii\ regions \citep{Mierkiewicz:2006ky,Madsen:2006fw}. \citet{Howk:2012tk} identified similar trends in UV absorption line observations.

Like \citet{Reynolds:1973jo}, \citet{Wood:1999bd} found that $5-20 \%$ of the \ha\ emission not directly associated with classical \hii\ regions could be scattered light from midplane \hii\ regions, not emitted {\it in situ} from ionized gas. However, the distinct spectral properties of the WIM combined with the evidence from pulsars that free electrons in the solar neighborhood extend to much larger heights than the OB stars are difficult to explain with a significantly larger scattered light contribution. By correlating $100 \mu \textrm{m}$ and \ha\ emission and extrapolating from the high-latitude dust cloud LDN~1780, \citet{Witt:2010gd} argued that the scattered light contribution could be up to $50 \%$ on sightlines with a large dust column and estimate that the most probable scattered light contribution is $0.1$~R, about $20 \%$ of the most probable high-latitude \ha\ intensity of $0.52$~R. They predict a large scattered light component and therefore low [{\sc Sii}] and [{\sc Nii}] in many sightlines in the southern Milky Way; we will test this hypothesis in new observations with the Wisconsin \ha\ Mapper, now located at Cerro Tololo in Chile.

\citet{Seon:2012ci} assert that the emission attributed to the WIM is dominated by scattered light. They have constructed models which reproduce the observed [{\sc Nii}]$\lambda 6584 / \ha$ and [{\sc Sii}]$/ \ha$ line ratios at high latitude through scattered light from late O and B star \hii\ regions. However, their model does not provide any explanation for the similarity of the line ratios in diffuse emission from the solar neighborhood at high latitudes and at altitudes well above the majority of the late OB stars ($|z| \gtrsim 0.5 \textrm{ kpc}$) in the Perseus Arm and edge-on galaxies. They note that all O9 and later stars only produce enough ionizing photons to account for half of the diffuse \ha, so this scattered light can at most be one of a number of contributors to the optical emission. \citet{Seon:2012ci} also argue that the \ha\ from the WIM may be overestimated due to contamination by \ha\ from stars. However, this cannot explain the observation that [{\sc Nii}] and [{\sc Sii}] are bright while [{\sc Oiii}] is faint (relative to \ha) in the WIM. Moreover, they argue that some of the [{\sc Sii}] emission attributed to the WIM may be emitted in neutral gas. While this is plausible in principle, [{\sc Oi}]$\lambda 6300$ observations indicate that the H ionization fraction within the optically emitting gas is $\gtrsim 0.9$ \citep{Reynolds:1998ji}.

\bibliography{papers_bibtex}

\end{document}